\documentstyle[11pt,moriond,epsfig]{article}
\begin{document}\vspace*{4cm}
\title{Constraints on Cosmological Parameters from Existing CMB Data}
\author{J.G. Bartlett, A. Blanchard, M. Le Dour, M. Douspis \& D. 
	Barbosa$^t$}
\address{Observatoire de Strasbourg, 11, rue de l'Universit\'e, 
	67000 Strasbourg, France\\
	$^t$ present address: Astronomy Centre, University of Sussex,
	Falmer, Brighton BN2 1TN, UK}
\maketitle
\abstracts{We use current cosmic microwave background data
to constrain cosmological parameters.  The results are presented
as confidence regions in the ($\Omega$--$H_o$) -- plane for both
open and flat geometries, and are based on a $\chi^2$ 
minimization to the power spectrum.  
Although such an approach does not strictly apply to power 
spectrum estimates, the results should nonetheless be indicative.  
We find that purely open, low--$\Omega$ models are strongly disfavored, due
to the position of the ``Doppler Peak'' at high $l$; flat
models are less strongly constrained.  The parameter space
explored is the largest to date, covering $\Omega$, $H_o$, 
$n$, $Q$ and $\Omega_B h^2$.}

\section{Introduction}

	It is already possible, with existing data on the 
fluctuations of the cosmic microwave background (CMB), to
constrain some cosmological parameters (Rocha \& Hancock 1996;
Bond \& Jaffe 1996; Lineweaver et al. 1997).  Of course, the actual data 
is far from providing the kind of constraints eagerly
awaited from the next generation of experiments 
(Jungman et al. 1996; Bond et al. 1997), but
there are nevertheless more than 10 different experiments
detecting fluctuations over a large range of angular scales.
To place our present work in context, consider
the general problem of analyzing the results (say, a map) 
of a CMB experiment.  If the
true sky fluctuations and, as is fortunately often the case,
the instrumental noise are gaussian, then
the probability of obtaining any particular set of $N$
pixel values is a multivariate gaussian in 
the pixel vector $\vec{d}$ (a one--dimensional listing of the
possible pixel values):
\begin{displaymath}
{\cal L}(\vec{p}) = {\cal P}(\vec{d}/\vec{p}) = 
	\frac{1}{(2\pi)^{N/2} |C|^{1/2}} e^{-(\vec{d})^t \cdot C^{-1} 
	\cdot (\vec{d})/2}
\end{displaymath}  
The cosmological model is completely
embodied (as well as the noise) in the covariance matrix
of the pixels, $C \equiv C^T(\vec{p}) + C(noise)$, where we denote
the underlying model parameters by the vector $\vec{p}$. 
Thus, we have a likelihood function for $\vec{p}$, given
a particular $\vec{d}$, i.e., a map of the sky.  It is 
important to note that even if the pixels are  
gaussian, the likelihood function ${\cal L}(\vec{p})$ is
NOT so, in general, because the parameters enter into the
{\em covariance} matrix and not as linear combinations
of the pixel values themselves.  Maximization of this
likelihood function is one way to estimate cosmological
parameters.

	Another approach is to work in Fourier space 
with estimates of the power spectrum, i.e., the $C_l$.
These are nothing more than the variances of the individual
spherical harmonic coefficients of
the temperature fluctuations.  We are really doing the
same thing as before, just working with 
the covariances of Fourier coefficients instead of the
measured pixel values.  A practical reason for using
this method is that band--power
estimates and their uncertainties are readily available in the 
literature, this being the most common way of concisely 
reporting the results of an experiment.

	In the present contribution, we assemble various 
band--power estimates (e.g., $C_l$ intergrated over
experimental window functions) and constrain certain
cosmological parameters by fitting to these power
spectrum data points.  We must mention several 
caveats applying to the results given here: 
Firstly, the data set is certainly not
homogeneous, if for no other reason than for the different
ways that the so--called cosmic variance has been 
estimated and included in the quoted error bars.
Secondly, we have not yet fully accounted for the
experimental window functions.  A third important point is
that the $C_l$ are NOT gaussian random variables, unlike
the individual pixel values (or Fourier coefficients).  
This is clear -- the $C_l$ represent the variance of
gaussian random variables, and thus are themselves distributed 
according to a $\chi^2$ distribution.
We have, for the work presented in this contribution,
effectively assumed gaussianity in the sense that we
apply a $\chi^2$ minimization for the fitting.  In principle,
this could lead to biased best--estimates and incorrect
confidence intervals.  We are currently 
working on improving these aspects of our calculations.
The results presented here should therefore be taken
as indicative, but perhaps relatively good indications,
all the same, given the current status of affairs (Jaffe \& Bond 1997).\\

\begin{tabular}{|c||c|c||c|c|}
\multicolumn{5}{c}{\bf\large \hspace*{1cm} Open Models\hspace*{4.5cm} Flat Models} \\
\hline\hline
parameter & range & step & range & step \\
\hline\hline
$\Omega$ 	& 0 -- 0.95	 	& 0.05	& 0 -- 1	& 0.05 \\
$H_o$		& 15 -- 100 km/s/Mpc 	& 5 km/s/Mpc & 15 -- 100 km/s/Mpc 	& 5 km/s/Mpc \\
$n$		& 0.5 -- 1.5		& 0.03	& 0.55 -- 1.45	& 0.09 \\
$Q$		& 14 -- 20 $\mu$K	& 1 $\mu$K & 14 -- 20 $\mu$K 	& 1 $\mu$K\ \\
$\Omega_B h^2$	&    0.015		& ---- & 0.006 -- 0.030	& 0.002  \\
\hline\hline
\end{tabular}

\section{The models}

	We consider two broad class of inflation--based models (e.g.,
gaussian fluctuations): Open and flat with a non--zero cosmological
constant.  For the {\bf open} models, we explored the four--dimensional
parameter space of {$\Omega,H_o,n,Q$}, where $n$ is the spectral index
of an assumed power--law power spectrum and $Q$ is the normalization
expressed as the CMB quadrupole; the baryon 
density was fixed at its nucleosythesis predicted value of
$\Omega_b h^2 = 0.015$.  For the {\bf flat} models, we varied
the baryon density in addition to these four parameters, thereby
exploring a five--dimensional space.  This is all summarized in 
the Table, where we also provide the individual step sizes on
each parameter and the range covered.  The calculations were
performed with CMBFAST (Seljak \& Zaldarriaga 1996).
To cover the parameter space indicated, the slower 
open model calculations required a couple of months of 
uninterrupted computing time. 

\section{Results}

 	Our results in the $(\Omega,H_o)$--plane for both
types of model are shown in Figure 1.  The light solid contours are defined 
by $\chi^2_{\rm min} + 2.3$ and $+6.17$, which would 
correspond to $1\sigma$ and $2\sigma$ limits in this plane
for a gaussian likelihood function (of the {\em parameters}),
which, as we have mentioned, is not really the case.  Nevertheless,
this gives some idea of the region favored by the
data: in general, we find that it corresponds to high $\Omega$ and
low values of $H_o$.  The dashed lines 
show the contours whose projection onto the axes give the 
confidence range on each parameter individually.  Finally, 
for comparison, the dark 
solid lines (outer contours) are contours of constant 
{\em goodness--of--fit} for, if all were normal, 68\% and 95\% 
probability.  

\begin{figure}
\centerline{
\psfig{figure=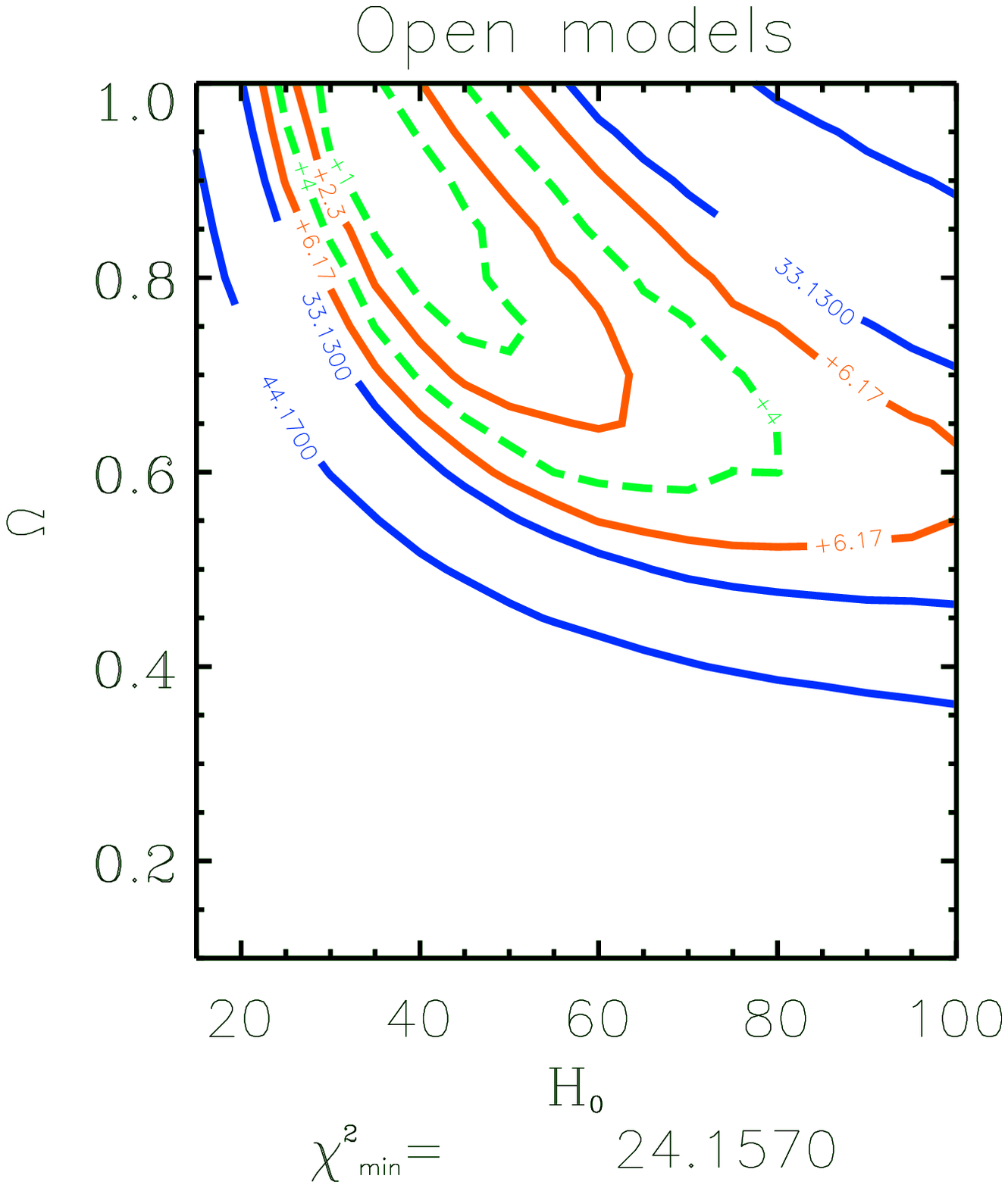,height=2.5in}
\psfig{figure=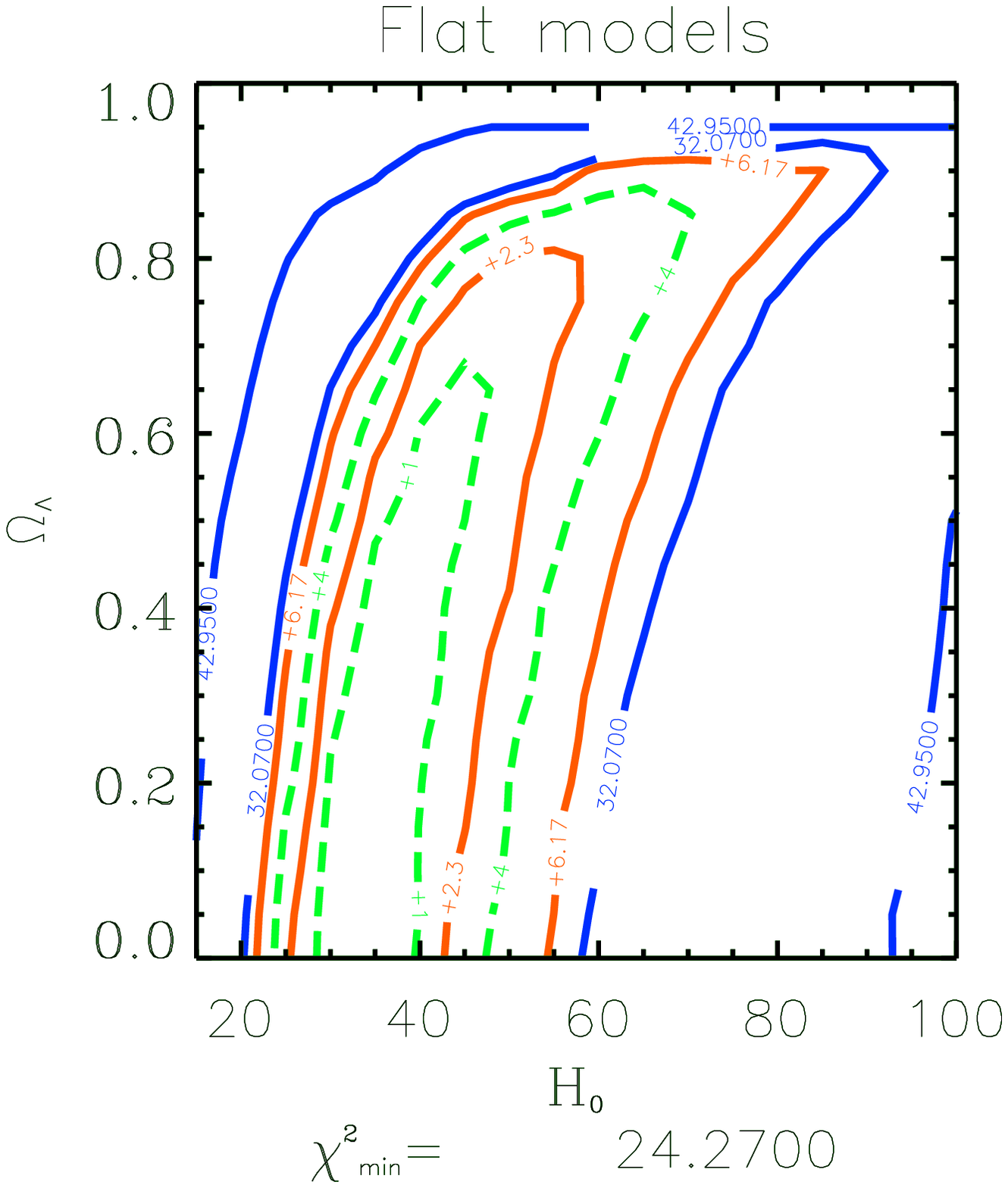,height=2.5in}}
\caption{LEFT: Contours in the $(\Omega,H_o)$ -- plane
for the {\bf Open} models.  The light
colored solid lines define confidence contours corresponding,
if everything were gaussian, to 68\% ($\chi_{min}^2+2.3$) and
95\% ($\chi_{min}^2+6.17$).  The dashed lines are confidence
ellipses whose projection onto the axes provides the estimated
range on each individual parameter.  The darker solid lines
are contours of constant goodness--of--fit at 68\% and
95\% probability, respectively, if the statistics were 
gaussian.  
RIGHT: Contours in the $(\Omega_v,H_o)$ -- plane 
for the {\bf Flat} models, where $\Omega_v = 1-\Omega$ is the
vacuum density.  The contours are defined as in the left--hand
panel.}
\end{figure} 

\begin{figure}
\centerline{
\psfig{figure=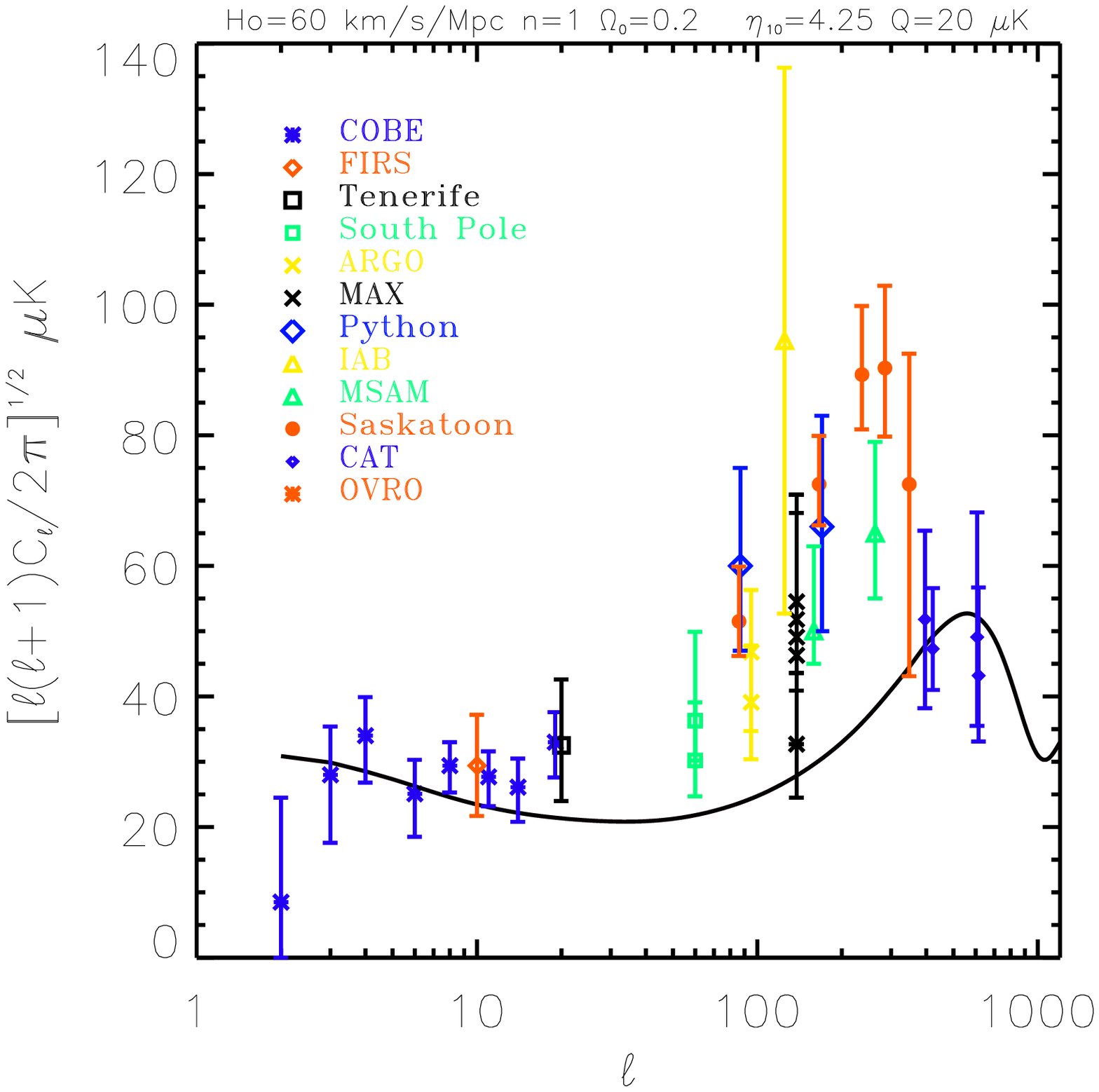,height=2.5in}
\psfig{figure=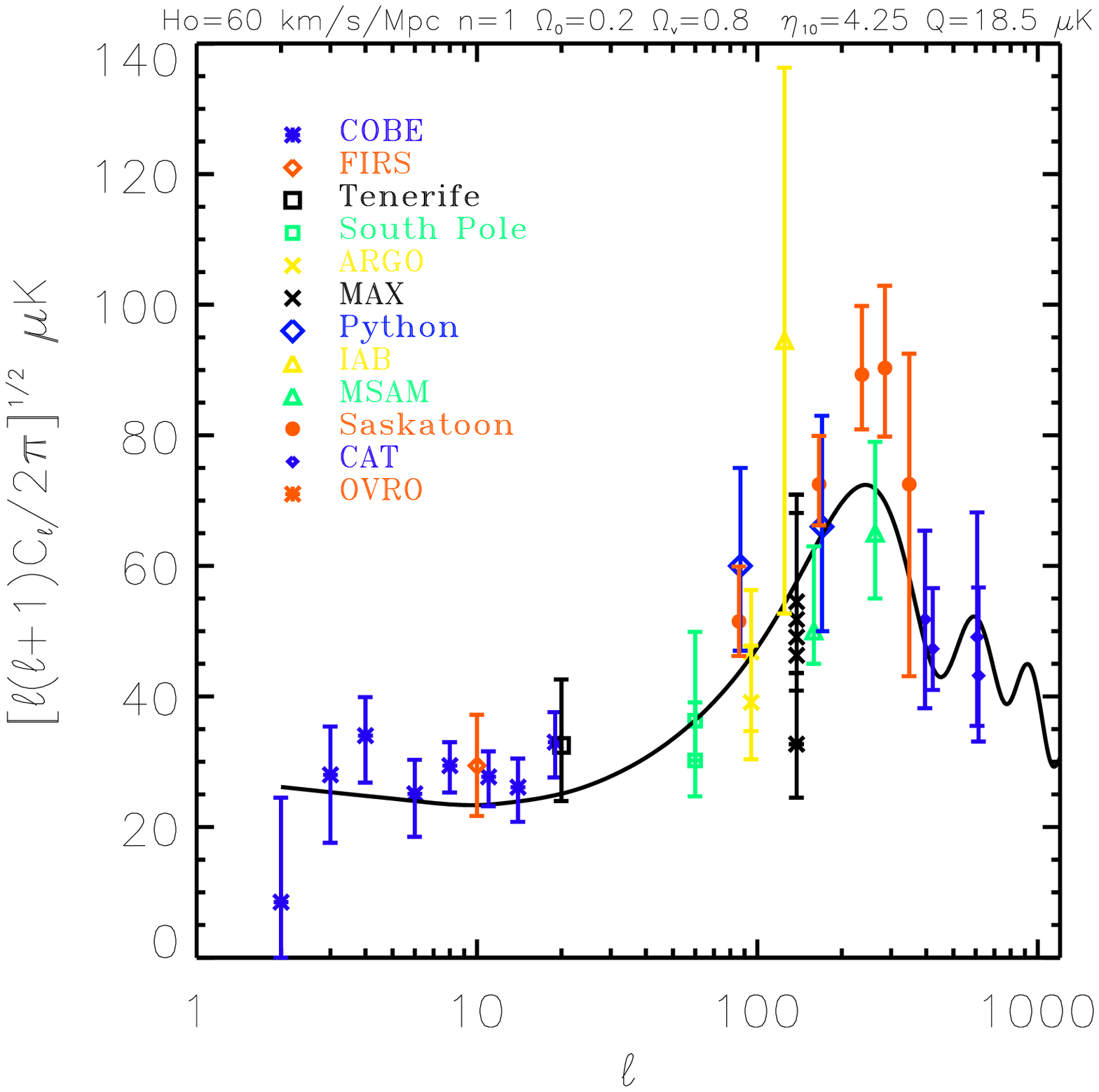,height=2.5in}}
\caption{LEFT: Power spectrum of an {\bf Open} model, with the
parameters given at the top of the figure, compared to the
data.  The relation between $\eta_{10}$ and the baryon density
is: $\Omega_B h^2 = 0.0036 \eta_{10}$, for a present--day CMB
temperature of $T=2.726$ K. 
RIGHT: Power spectrum of the corresponding {\bf Flat} model.}
\end{figure}

	The most interesting aspect of these results concerns 
the open models.  We see from Figure 1 that the contours 
place a lower limit on $\Omega$; values as low 
as $\Omega=0.2$ are strongly disfavored for any $H_o$.
The reason for this 
is clear from the left--hand--side of Figure 2.
The ``Doppler Peak'' is displaced too far to the right, due
to the focusing of light rays in an open geometry
(Hu \& White 1996).  Thus 
we find that a ``comfortable'' model with $\Omega=0.2$ and
$H_o=60$ km/s/Mpc has {\em very serious difficulty} with the 
CMB data.  

	How about the cosmological constant?  Here, the constraints
on $\Omega$ are much less stringent, as can be seen from the contours
shown in the right--hand panel of Figure 1.  In contrast to 
the purely open model with $\Omega=0.2$, the corresponding flat 
model, with $\Lambda=0.8$, is acceptable and provides a good--looking
fit to the power spectrum (right--hand--side of Figure 2).

\section{Conclusions}

	It is a little early to draw detailed conclusions 
from the CMB concerning the cosmological parameters, but it is worth
noting that some conclusions are already possible with existing
data; and the next round of data releases will not be long in coming.
Despite the various caveats of the method used here, it seems
that one result should remain rather robust - that low--$\Omega$,
purely open models are strongly disfavored, for the 
clear and simple reason that the ``Doppler Peak'' is just not
in the correct place due to the large angular distance to the
surface of last scattering in such models.

\section{Acknowledgements}  

	D.B acknowledges support by a Ph.D. Praxix XXI grant attributed
by JNICT/FDC, Portugal.


\section*{References}
\begin{thebibliography}{99}
\small
\bibitem{} Bond J.R., Efstathiou G. \& Tegmark M. 1997, MNRAS 291, L33
\bibitem{} Bond J.R. \& Jaffe A.H. 1996, in ``Microwave Background
	Anisotropies'' (proceedings of the XVIth Moriond 
	Astrophysics meeting),
	F.R. Bouchet, R. Gispert et al. (Eds), Editions Fronti\`eres
	(Gif--sur--Yvette) p. 197
\bibitem{} Hu W. \& White M. 1996, ApJ 471, 30
\bibitem{} Jaffe A.H. \& Bond J.R. 1997, proceedings of the
	18th Texas Symposium, astro-ph/9702109
\bibitem{} Jungman G., Kamionkowski M., Kosowsky A. \& Spergel D.
	1996, PhyRev D54, 1332 
\bibitem{} Lineweaver C.H., Barbosa D., Blanchard A. \& Bartlett J.G.
	1997, A\&A 322, 365
\bibitem{} Rocha G. \& Hancock S. 1996, in ``Microwave Background
	Anisotropies'' (proceedings of the XVIth Moriond 
	Astrophysics meeting),
	F.R. Bouchet, R. Gispert et al. (Eds), Editions Fronti\`eres
	(Gif--sur--Yvette) p. 189
\bibitem{} Seljak U. \& Zaldarriaga M. 1996, ApJ 469, 437
\end{thebibliography}
\end{document}